\documentclass{ws-ijmpb}

\usepackage{graphicx}
\usepackage{verbatim}
\usepackage{color,hyperref}
\usepackage{amsfonts}

\newcommand{\beq}{\begin{equation}}
\newcommand{\eeq}{\end{equation}}
\newcommand{\beqa}{\begin{eqnarray}}
\newcommand{\eeqa}{\end{eqnarray}}

\begin{document}

\title{Renormalization of the superfluid density \\
in the two-dimensional BCS-BEC crossover}

\author{G. Bighin}
\address{IST Austria (Institute of Science and Technology Austria), 
Am Campus 1, 3400 Klosterneuburg, Austria}

\author{L. Salasnich}
\address{Dipartimento di Fisica e Astronomia ``Galileo Galilei'' 
and CNISM, Universit\`a di Padova, \\ via Marzolo 8, 35131 Padova, Italy \\
CNR-INO, via Nello Carrara, 1 - 50019 Sesto Fiorentino, Italy}

\date{\today}

\maketitle

\begin{history}
\received{Day Month Year}
\revised{Day Month Year}
\end{history}

\begin{abstract}
We analyze the theoretical derivation of the beyond-mean-field equation 
of state for a two-dimensional gas of dilute, ultracold alkali-metal atoms 
in the Bardeen-Cooper-Schrieffer (BCS) to Bose-Einstein condensate (BEC) 
crossover. We show that at zero temperature our theory -- considering Gaussian
fluctuations on top of the mean-field equation of state -- is in 
very good agreement with experimental data. Subsequently, we investigate 
the superfluid density at finite temperature and 
its renormalization due to the proliferation of vortex-antivortex pairs. 
By doing so, we determine the Berezinskii-Kosterlitz-Thouless (BKT) critical 
temperature -- at which the renormalized superfluid density jumps to zero --
as a function of the inter-atomic potential strength. We find that the 
Nelson-Kosterlitz criterion overestimates the BKT temperature 
with respect to the renormalization group equations, 
this effect being particularly relevant in the intermediate regime of 
the crossover. 
\end{abstract}


\maketitle

\section{Introduction} 
 
In 2004 the three-dimensional crossover between 
the Bardeen-Cooper-Schrieffer (BCS) regime 
of weakly attractive fermions to the Bose-Einstein 
condensate (BEC) regime of strongly-bound bosonic molecules has been 
realised using ultracold, two-component fermionic 
$^{40}$K or $^6$Li atoms\cite{regal,zwierlein,kinast}.
The crossover is obtained using a Fano-Feshbach resonance to tune 
the s-wave scattering length $a_F$ of the 
inter-atomic potential. Recently, the two-dimensional 
BEC-BEC crossover has been 
achieved experimentally\cite{makhalov,ries,fenech,boettcher} 
using a two-component fermionic $^6$Li atoms confined in a (quasi-) 
two-dimensional geometry.
The properties of two-dimensional fermions are quite different with respect 
to their three-dimensional counterpart,
in particular, in two dimensions, attractive fermions always form a 
bound-state with energy 
$\epsilon_B \simeq \hbar^2/(m a_F^2)$, 
where $a_F$ is the two-dimensional s-wave scattering length. 
The fermionic single-particle spectrum is given by 
\beq 
E_{sp}(k) = \sqrt{\left({\hbar^2k^2\over 2m}-\mu\right)^2+\Delta_0^2} \; ,
\label{eq:esp0}
\eeq
where $\Delta_0$ is the energy gap and $\mu$ is the chemical potential: 
$\mu > 0$ corresponds to the BCS regime while $\mu<0$ corresponds to the 
BEC regime. Moreover, in the deep BEC regime $\mu \to -\epsilon_B/2$. 

\section{Two-dimensional equation of state}

To study the two-dimensional BCS-BEC crossover we adopt the formalism 
of functional integration\cite{nagaosa}. The partition function ${\cal Z}$
of a uniform system of ultracold, dilute, interacting spin $1/2$ fermions
at temperature $T$, in a two-dimensional volume $L^2$, 
with chemical potential $\mu$ reads
\beq 
{\cal Z} = \int {\cal D}[\psi_{s},\bar{\psi}_{s}] 
\ \exp{\left\{ -{S\over \hbar} \right\} } \; , 
\eeq
where the complex Grassmann field 
$\psi_{s} ({\bf r},\tau )$, $\bar{\psi_{s}} ({\bf r},\tau )$ describes 
the fermions, $\beta \equiv 1/(k_B T)$ with $k_B$  Boltzmann's constant and
\beq 
S = \int_0^{\hbar\beta} 
d\tau \int_{L^2} d^2{\bf r} \ {\cal L}
\eeq
is the Euclidean action functional with Lagrangian density
\beq 
{\cal L} = \bar{\psi}_{s} \left[ \hbar \partial_{\tau} 
- \frac{\hbar^2}{2m}\nabla^2 - \mu \right] \psi_{s} 
+ { g} \, \bar{\psi}_{\uparrow} \, \bar{\psi}_{\downarrow} 
\, \psi_{\downarrow} \, \psi_{\uparrow}
\label{eq:ham0}
\eeq
${g}$ being the attractive strength (${ g}<0$) 
of the s-wave coupling. 

Through the usual Hubbard-Stratonovich transformation 
the Lagrangian density ${\cal L}$ -- quartic in the fermionic fields --
can be rewritten  as a quadratic form by introducing the
auxiliary complex scalar field $\Delta({\bf r},\tau)$. 
After doing so, the effective Euclidean Lagrangian density reads 
\beq 
{\cal L}_e =
\bar{\psi}_{s} \left[  \hbar \partial_{\tau} 
- {\hbar^2\over 2m}\nabla^2 - \mu \right] \psi_{s} 
+ \bar{\Delta} \, \psi_{\downarrow} \, \psi_{\uparrow} 
+ \Delta \bar{\psi}_{\uparrow} \, \bar{\psi}_{\downarrow} 
- {|\Delta|^2\over { g}} \; . 
\label{ltilde}
\eeq 

We investigate the effect of 
fluctuations of the pairing field $\Delta({\bf r},t)$ around its
mean-field value $\Delta_0$ which may be taken to be real. 
For this reason we set 
\beq 
\Delta({\bf r},\tau) = \Delta_0 +\eta({\bf r},\tau) \; , 
\label{polar}
\eeq
where $\eta({\bf r},\tau)$ is the complex field describing
pairing fluctuations. 
In particular, we are interested in the grand potential $\Omega$, 
given by 
\beq 
\Omega = - {1\over \beta} \ln{\left( {\cal Z} \right)} \simeq - {1\over \beta} 
\ln{\left( {\cal Z}_{mf} {\cal Z}_g \right)} = \Omega_{mf} + \Omega_{g} 
\; , 
\eeq 
where 
\beq 
{\cal Z}_{mf} = \int {\cal D}[\psi_{s},\bar{\psi}_{s}]\, 
\exp{\left\{ - {S_e(\psi_s, \bar{\psi_s}, 
\Delta_0) \over \hbar} \right\}} \;  
\eeq
is the mean-field partition function and 
\beq
{\cal Z}_g = \int {\cal D}[\psi_{s},\bar{\psi}_{s}]\, 
{\cal D}[\eta,\bar{\eta}] \ 
\exp{\left\{ - {S_g(\psi_s, \bar{\psi_s},
\eta,\bar{\eta},\Delta_0) \over \hbar} \right\}} 
\eeq
is the partition function of Gaussian pairing fluctuations.  
After functional integration over quadratic fields, 
one finds that the mean-field grand potential reads\cite{altland}
\beq
\Omega_{mf} = - {\Delta_0^2\over { g}} L^2 
+ \sum_{\bf k} \left( {\hbar^2k^2\over 2m} - \mu - E_{sp}({\bf k}) 
- {2\over \beta } \ln{(1+e^{-\beta\, E_{sp}({\bf k})})} 
\right) \; , 
\eeq
where $E_{sp}({\bf k})$ is the spectrum of fermionic single-particle 
excitations, as defined in Eq. (\ref{eq:esp0}).
On the other hand, the Gaussian-level grand potential is given by 
\beq
\Omega_{g} = {1\over 2\beta} \sum_{Q} \ln{\mbox{det}({\bf M}(Q))} \; ,  
\label{goduria}
\eeq
where ${\bf M}(Q)$ is the { inverse propagator of Gaussian 
fluctuations of pairs} and $Q=({\bf q},i\Omega_{m})$ is the 
$(2+1)$-dimensional 
wavevector with $\Omega_{m}=2\pi m/\beta$ the Matsubara frequencies 
and ${\bf q}$ the two-dimensional wavevector\cite{diener}.

The sum over Matsubara frequencies is quite complicated and it 
does not give a simple expression. An approximate 
formula\cite{taylor} is 
\beq
\Omega_{g} \simeq {1\over 2} 
\sum_{\bf q} E_{col}({\bf q}) + {1\over \beta }
\sum_{\bf q} \ln{(1- e^{-\beta\, E_{col}({\bf q})})} \; ,   
\eeq
where 
\beq 
E_{col}({\bf q}) = \hbar \ \omega({\bf q}) 
\eeq
is the spectrum of bosonic collective excitations with $\omega({\bf q})$ 
derived from 
\beq 
\mbox{det}({\bf M}({\bf q},\omega)) = 0 \; . 
\eeq
Notice that very recently a comprehensive experimental study of 
fermionic and bosonic elementary excitations in a homogeneous 
3D strongly interacting Fermi gas through the BCS-BEC crossover has been 
performed using two-photon Bragg spectroscopy.\cite{vale}

In our approach (Gaussian pair fluctuation theory\cite{hu}), 
the grand potential is given by
\beq 
\Omega(\mu,L^2,T,\Delta_0) = \Omega_{mf}(\mu,L^2,T,\Delta_0) + 
\Omega_g(\mu,L^2,T,\Delta_0) \; , 
\eeq
and the energy gap $\Delta_0$ is obtained from the (mean-field) gap equation 
\beq 
{\partial \Omega_{mf}(\mu,L^2,T,\Delta_0) \over \partial \Delta_0} = 0 \; . 
\eeq
The number density $n$ is instead obtained from the number equation 
\beq 
n = - {1\over L^2} {\partial \Omega(\mu,L^2,T,\Delta_0(\mu,T)) 
\over \partial \mu} 
\eeq
taking into account the gap equation, i.e. that $\Delta_0$ is a function 
$\Delta_0(\mu,T)$ of $\mu$ and $T$.
Notice that the Nozi\`eres-Schmitt-Rink 
approach\cite{noziers} is quite similar but neglects, in the number equation,
that $\Delta_0$ depends on $\mu$ .

\section{Zero-temperature results} 

In the analysis of the two-dimensional attractive Fermi gas
one must remember that, as opposed to the three-dimensional case, 
two-dimensional realistic interatomic attractive potentials always have 
a bound state. In particular\cite{mora}, 
the binding energy ${ \epsilon_B}>0$ of two fermions can be written 
in terms of the positive two-dimensional fermionic 
scattering length ${ a_F}$ as 
\beq 
{ \epsilon_B}= {4\over e^{2\gamma}}{\hbar^2\over 
m { a_F}^2} \; ,   
\label{eb-af}
\eeq
where $\gamma=0.577...$ is the Euler-Mascheroni constant. Moreover, 
the attractive s-wave interaction strength $g$ appearing 
in Eq. (\ref{eq:ham0}) is related to the 
binding energy ${ \epsilon_B}>0$ of a fermion pair in vacuum by 
the expression\cite{randeria,sala2007}
\beq 
- \frac{1}{ g} 
= \frac{1}{2L^2} \sum_{\bf k} \frac{1}{{\hbar^2k^2\over 2m} + 
\frac{1}{2} { \epsilon_B}} \; . 
\label{g-eb}
\eeq 

\begin{figure}
\begin{center}
\includegraphics[width=8cm,clip=]{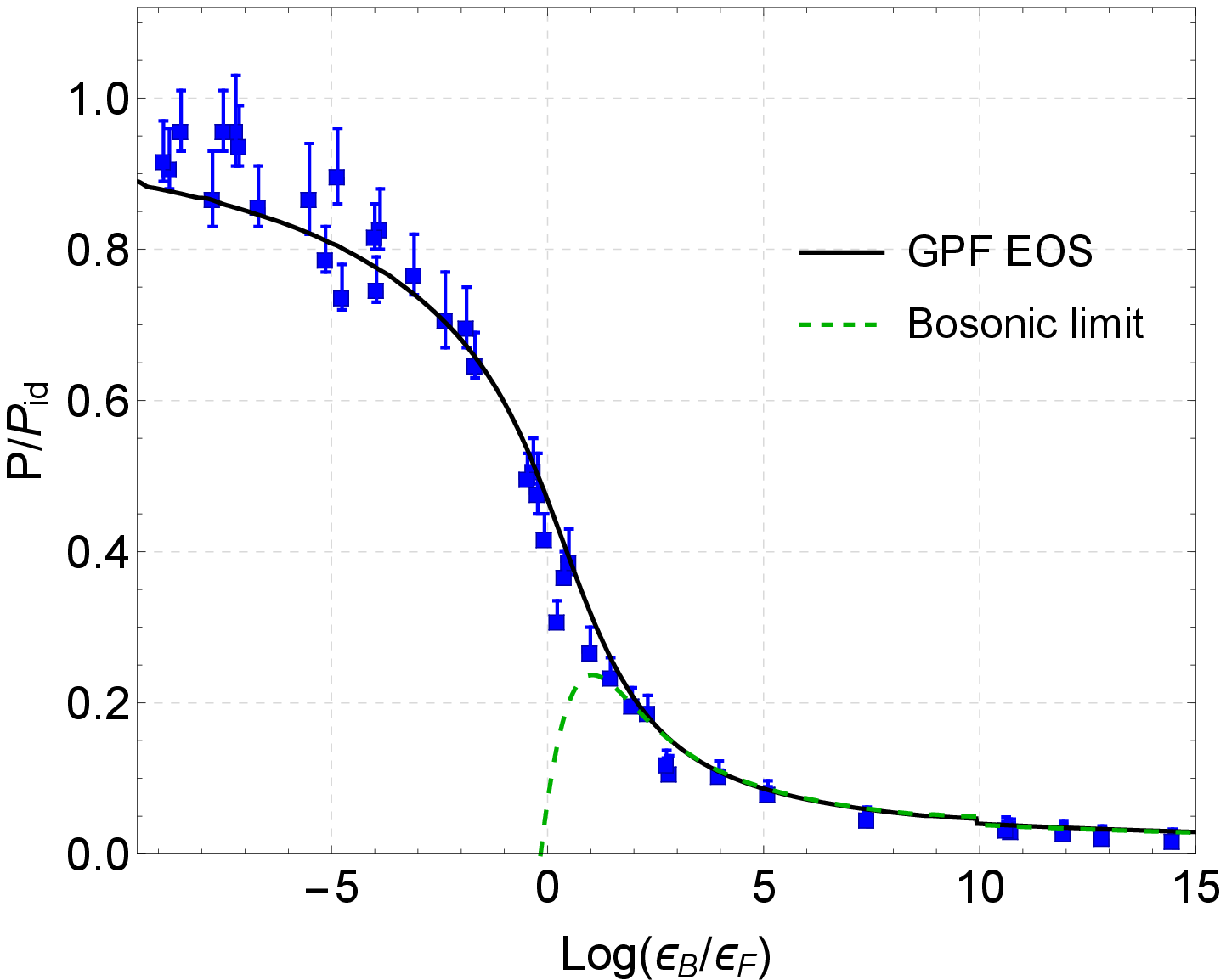}
\end{center}
{\bf Fig. 1}. Scaled pressure $P/P_{id}$ vs scaled binding 
energy ${ \epsilon_B}/\epsilon_F$. Notice that $P=-\Omega/L^2$ 
and $P_{id}$ is the pressure of the ideal two-dimensional Fermi gas. 
Filled squares with error bars: experimental data of Makhalov {\it et al.} 
\cite{makhalov}. Solid black line: the regularized Gaussian pair (GP) 
theory\cite{bighin2016}. Dashed green line: Popov equation of state, 
Eq. (\ref{iomeio}), of bosons with mass $m_B=2m$.
\end{figure} 

At zero temperature, including Gaussian fluctuations, the pressure is  
\beq 
P = - {\Omega\over L^2} = {m L^2\over 2\pi \hbar^2} 
(\mu + {1\over 2} { \epsilon_B} )^2  + P_g(\mu,L^2,T=0) \; ,  
\eeq 
with  
\beq 
P_g(\mu,L^2,T=0) = - {1\over 2} \sum_{\bf q} E_{col}({\bf q}) \; .  
\eeq 
In the full two-dimensional BCS-BEC crossover, from the { regularized} 
version of Eq. (\ref{goduria}), we obtain numerically the zero-temperature  
pressure\cite{bighin2016} (see also Ref. \cite{he}). 
The results are shown in Fig. 1, 
where the agreement with the experimental data\cite{makhalov} 
is very satisfying.

In the deep-BEC regime the chemical potential $\mu$ is negative and large 
in modulus. The energy of bosonic collective excitations becomes 
\beq
E_{col}({\bf q}) = \sqrt{ {\hbar^2q^2\over 2m} 
\left( \lambda {\hbar^2q^2\over 2m} + 2 m c_s^2 \right) }
\eeq
with $\lambda=1/4$ and 
$m c_s^2=\mu + { \epsilon_B}/2$. Moreover, 
the corresponding regularized pressure 
-- which can be obtained by means of dimensional 
regularization\cite{toigo2015,sala2017} -- reads 
\beq 
P = {m \over 64\pi\hbar^2}  
(\mu + {1\over 2}{ \epsilon_B})^2 \ 
\ln{\left({{ \epsilon_B}\over 
2 (\mu + {1\over 2}{ \epsilon_B}) } \right)} \; .  
\label{iomeio}
\eeq 
This is exactly the Popov equation of state of two-dimensional Bose gas 
with chemical potential $\mu_B = 2(\mu + { \epsilon_B}/2)$ and 
boson mass $m_B=2m$. 
In this way we have identified the two-dimensional scattering 
length ${ a_B}$ of composite bosons as
\beq 
a_B = {1\over 2^{1/2}e^{1/4}} \ a_F \; .
\eeq
The value ${ a_B}/{ a_F}= 1/(2^{1/2}e^{1/4}) \simeq 0.551$ 
is in full agreement with the value ${ a_B}/{ a_F}=0.55(4)$ 
obtained by Monte Carlo calculations\cite{bertaina}.

\section{Quantized vortices and superfluid density} 

In Section II we have written the pairing field through 
Eq. (\ref{polar}). A different parametrisation\cite{marchetti} is provided by
\beq 
\Delta({\bf r},\tau) = \left( \Delta_0 + \sigma({\bf r},\tau) \right) \ 
e^{i \theta({\bf r},\tau)} \; , 
\eeq
where $\sigma({\bf r},\tau)$ is the real field of 
amplitude fluctuations and $\theta({\bf r},\tau)$ is the 
angular field of { phase fluctuations}. 
However, Taylor-expanding the exponential of the phase, one has 
\beq 
\left( \Delta_0 + \sigma({\bf r},\tau) \right) \ 
e^{i \theta({\bf r},\tau)} = \Delta_0 + \sigma({\bf r},\tau) 
+ i \ \Delta_0  \ \theta({\bf r},\tau )+ ... \; .  
\eeq
Thus, at the Gaussian level, we can write 
\beq 
\eta({\bf r},\tau) =  \sigma({\bf r},\tau) + i \ 
\Delta_0  \ \theta({\bf r},\tau ) \; . 
\eeq
After functional integration over $\sigma({\bf r},\tau)$, the 
Gaussian action becomes 
\beq 
S_g = \int_0^{\hbar\beta} 
d\tau \int_{L^2} d^2{\bf r} \ \left\{ {J \over 2}  
\left( \nabla \theta \right)^2 + 
{\chi \over 2} \left( {\partial\theta\over \partial \tau}\right)^2 \right\} 
\label{enda1}
\eeq
where $J$ is the { phase stiffness} and 
$\chi$ is the { compressibility}. 
This is the quantum action of the 2D continuous XY model.\cite{altland} 
The superfluid density is related to 
the phase stiffness $J$ by the simple formula 
\beq 
n_s = \frac{4m}{\hbar^2} J \; . 
\eeq

At the Gaussian level $J$ depends only on fermionic single-particle 
excitations $E_{sp}(k)$.\cite{babaev} However, beyond the Gaussian 
level also bosonic collective 
excitations $E_{col}(q)$ contribute\cite{benfatto}.
Thus, we assume the following Landau-type formula 
\beq
n_s(T) = n - \beta \int
\frac{\mathrm{d}^2 k}{(2 \pi)^2} k^2 \frac{e^{\beta E_{sp}(k)}}
{(e^{\beta E_{sp}(k)} + 1)^2}
- \frac{\beta}{2} \int \frac{\mathrm{d}^2 q}{(2 \pi)^2} q^2
\frac{e^{\beta E_{col}(q)}}{(e^{\beta E_{col}(q)} - 1)^2} \;  
\eeq
where both fermionic and bosonic elementary excitations are included. 

\begin{figure}
\begin{center}
\includegraphics[width=9cm,clip=]{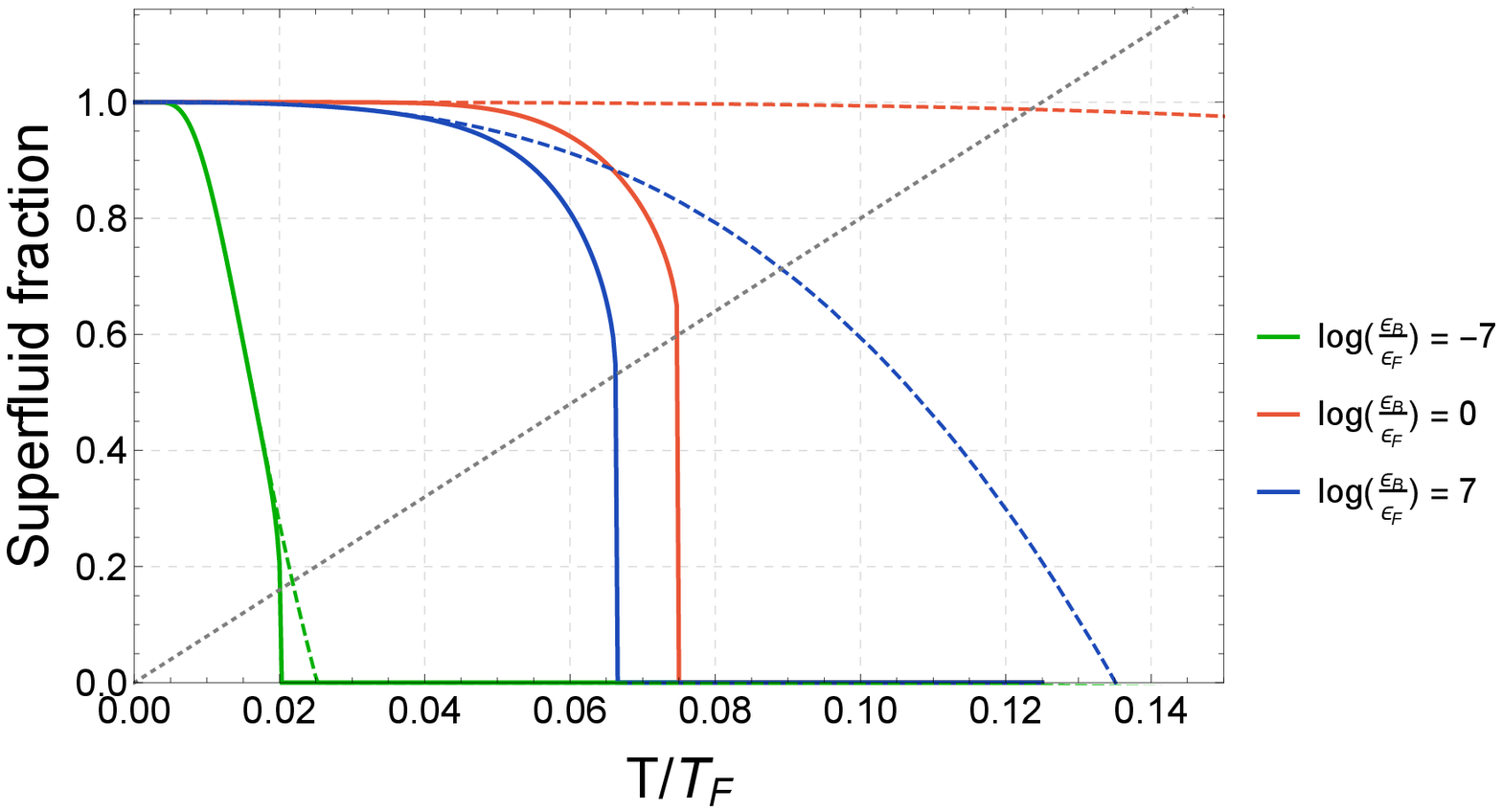}
\end{center}
{\bf Fig. 2}. Superfluid fraction $n_s/n$ vs scaled temperature $T/T_F$ 
in the two-dimensional BEC-BEC crossover.\cite{bighin2017} Solid lines: 
renormalized 
superfluid density. Dashed lines: bare superfluid density. 
$T_F=\epsilon_F/k_B$ is the Fermi temperature. 
Gray dotted line: Nelson-Kosterlitz condition 
$k_BT =(\pi/2) J(T)= (\hbar^2\pi/(8m)) n_s(T)$. 
\end{figure} 

It is important to stress that the compactness of the 
phase angle $\theta({\bf r},t)$ implies that 
\beq 
\oint_{\cal C} {\boldsymbol \nabla} 
\theta({\bf r},t) \cdot d{\bf r} = 2\pi \sum_i q_i \; , 
\eeq
where $q_i$ is the integer number associated to quantized vortices  
($q_i>0$) and antivortices ($q_i <0$) encircled by ${\cal C}$. 
One can write\cite{altland} 
\beq 
{\boldsymbol \nabla} \theta({\bf r},t) = 
{\boldsymbol \nabla} \theta_0({\bf r},t) - {\boldsymbol \nabla} \wedge 
\left( {\bf u}_z \ \theta_v({\bf r}) \right) \; , 
\eeq
where ${\boldsymbol \nabla} \theta_0({\bf r},t)$ 
has zero circulation (no vortices) while 
$\theta_v({\bf r})$ encodes the contribution of 
quantized vortices and anti-vortices, namely 
\beq 
\theta_v({\bf r}) = \sum_i q_i 
\ln{\left({\frac{|{\bf r}-{\bf r}_i|}{\xi}}\right)} \; , 
\label{enda2}
\eeq
where ${\bf r}_i$ is the position of the i-th vortex and  
$\xi$ is the cutoff length defining the vortex core size, 
with $\mu_v$ its energy. From Eqs. (\ref{enda1}) and (\ref{enda2}) 
one finds that the attractive intraction potential of 
a vortex-antivortex pair (with $q_i=1$ and $q_j=-1$) 
is proportional to the phase stiffiness 
$J$ and is given by\cite{altland} 
\beq
V_v({\bf r}_i,{\bf r}_j) = - 2\pi \ J \ 
\ln{\left({\frac{|{\bf r}_i-{\bf r}_j|}{\xi}}\right)} \; .
\label{enda3}
\eeq

The analysis of Kosterlitz and Thouless\cite{kosterlitz} 
of the two-dimensional XY model shows that: 
\begin{itemize}
\item 
As the temperature $T$ increases vortices start to appear 
in vortex-antivortex pairs (mainly with $q=\pm 1$). 
\item
The pairs are bound at low temperature 
until, at the critical temperature $T_{BKT}$ of 
Berezinskii-Kosterlitz-Thouless,\cite{nagaosa,altland} 
an unbinding transition occurs above which a proliferation of 
free vortices and antivortices is predicted. 
\item 
The phase stiffness $J$ and the vortex energy $\mu_v$ are renormalized 
due the screening of other vortex-antivortex pairs 
on the interaction potential (\ref{enda3}). 
\item 
The renormalized superfluid density $n_{s,R}=J_R(4m/\hbar^2)$ 
decreases by increasing the temperature $T$ 
and jumps to zero at $T_{BKT}$. 
\item
The renormalized vortex energy $\mu_{v,R}$, that is the energy cost 
to produce a unbound vortex, is infinity for $T\leq T_{BKT}$. 
\end{itemize}

The renormalized phase stiffness $J_R$ is obtained from 
the bare one $J$ by solving the renormalization group (RG) 
equations\cite{nelson}
\beqa
\frac{\mathrm{d}}{\mathrm{d} \ell} K(\ell) &=& - 
4\pi^3 K(\ell)^2 y(\ell)^2 
\\
\frac{\mathrm{d}}{\mathrm{d} \ell} y(\ell) &=& 
\left( 2 - \pi K(\ell) \right) y(\ell) 
\eeqa
for the running variables $K(\ell)$ and $y(\ell)$, as a function of the 
adimensional scale $\ell$ subjected to the initial conditions
$K(\ell = 0) = J/\beta$ and $y(\ell = 0) = \exp(-\beta \mu_v)$, with 
$\mu_v = \pi^2 J/4$ the vortex energy\cite{zhang}.
The renormalized phase stiffness is then 
\beq 
J_R = \beta \ K(\ell = +\infty) \; , 
\eeq
and the corresponding renormalized superfluid density reads  
\beq 
n_{s,R} = \frac{4m}{\hbar^2} J_R \; . 
\eeq
In Fig. 2 we plot the superfluid fraction $n_s/n$ 
as a function of the temperature $T$ for three strengths 
of the BEC-BEC crossover. 
In the figure we report both the bare superfluid density (dashed lines) 
and the renormalized one (solid lines). Notice that 
the renormalized superfluid density satisfies the 
Nelson-Kosterlitz condition\cite{nelson} 
\beq 
k_BT_{BKT} ={\pi\over 2} J_R(T_{BKT}^-) = {\hbar^2\pi\over 8m} 
n_{s,R}(T_{BKT}^-) \; . 
\eeq 

\begin{figure}
\begin{center}
\includegraphics[width=9cm,clip=]{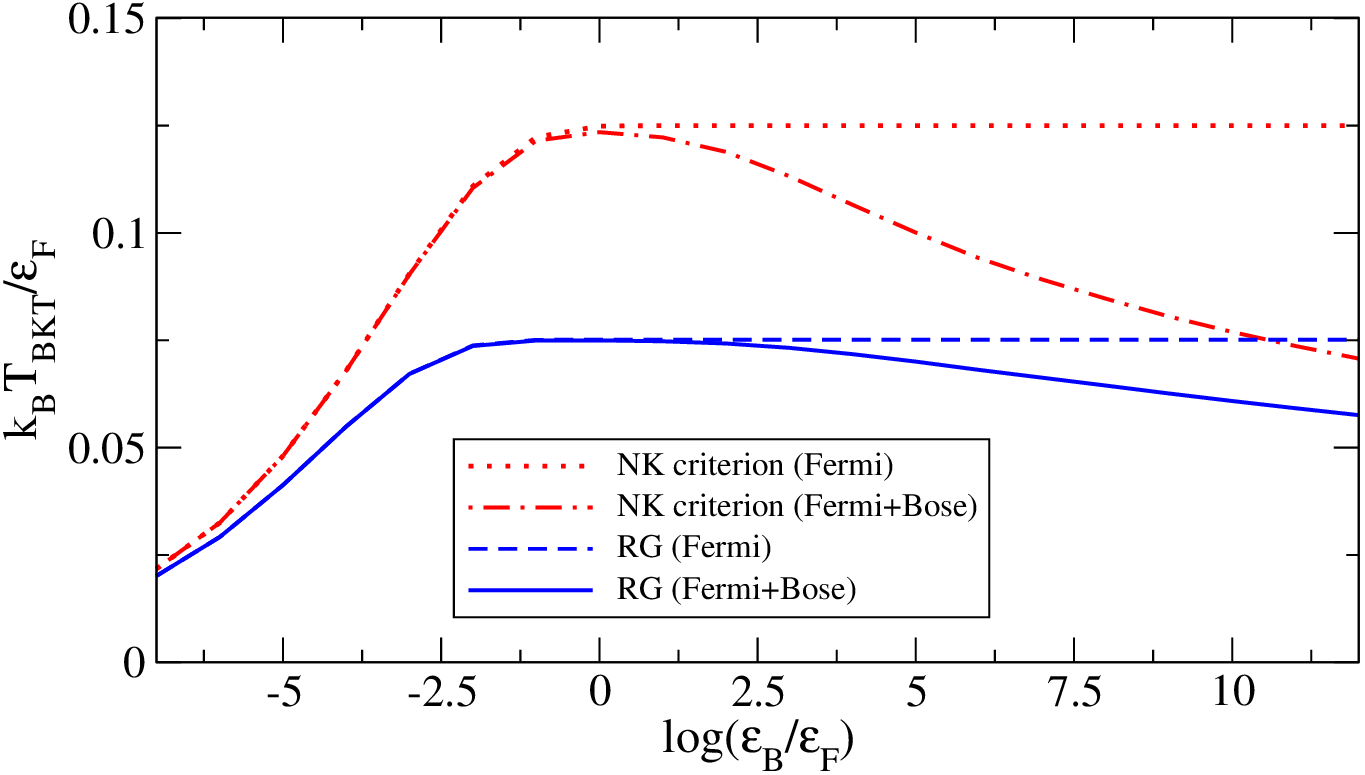}
\end{center}
{\bf Fig. 3}. Theoretical predictions for the 
Berezinskii-Kosterlitz-Thouless (BTK) 
critical temperature $T_{BKT}$. Red dot-dahsed and dashed lines obtained 
by using\cite{bighin2016} the 
Nelson-Kosterlitz (NK) condition on the bare superfluid density 
(NK criterion): 
$k_BT_{BKT}= (\hbar^2\pi/(8m)) n_s(T_{BKT})$. 
Blue sodlid and dashed lines obtained by solving\cite{bighin2017} 
the renormalization group (RG) equations. 
\end{figure} 

In Fig. 3 we report our theoretical predictions for the 
critical temperature $T_{BKT}$. Dot-dashed and dotted lines are 
obtained by using\cite{bighin2016} the 
Nelson-Kosterlitz condition with the bare superfluid density. 
This approach is called Nelson-Kosterlitz criterion. 
Solid and dashed lines are instead obtained by using\cite{bighin2017}
the Nelson-Kosterlitz condition on the renormalized superfluid density. 
The figure cleary shows that the inclusion of bosonic elementary 
excitations is crucial to get a reduction of $T_{BKT}$ the BEC regime. 
Moreover, the Nelson-Kosterlitz criterion, based on the 
Nelson-Kosterlitz condition with the bare superfluid density 
instead of the renormalized one, is not accurate 
in the middle of the crossover. 

\section{Conclusions} 

We have shown that, after regularization of Gaussian fluctuations 
(for a recent comprehensive review see Ref. \cite{toigo2016}), 
the beyond-mean-field theory of the two-dimensional BCS-BEC crossover 
is in very good agreement with (quasi) zero-temperature 
experimental data\cite{makhalov}. Moreover, in the BEC regime 
of the crossover the equation of state gives the 
correct logarithmic behavior characteristic of weakly-interacting 
repulsive bosons\cite{toigo2015}.
At finite temperature we have found that beyond-mean-field effects, 
as well the contribution from quantized vortices and antivortices, 
determine the properties of the two-dimensional BCS-BEC crossover. 
In particular, the inclusion of collective bosonic excitations is 
essential to get a reliable determination of the superfluid density 
and of Berezinskii-Kosterlitz-Thouless (BKT) critical temperature, 
across the whole crossover. Moreover, we have shown that, 
in the intermediate regime of the BCS-BEC crossover, 
the Nelson-Kosterlitz criterion strongly overestimate the critical 
temperature with respect to the results obtained through the 
renormalization group equations. 

\section*{Acknowledgements}

LS acknowledges partial support from the 2016 BIRD project 
``Superfluid properties of Fermi gases in optical potentials" of the 
University of Padova. The authors thank L. Dell'Anna, S. Klimin, 
P.A. Marchetti, J. Tempere, and F. Toigo for many enlightening discussions.

\end{document}